\documentclass[twocolumn,showpacs,preprintnumbers,amsmath,amssymb,aps]{revtex4}
\usepackage{graphicx}
\usepackage{dcolumn}
\usepackage{bm}

\begin{document}

\title{Optical frequency metrology with a Rb-stabilized ring-cavity resonator -- Study of cavity-dispersion errors}
 \author{Alok K. Singh}
 \author{Lal Muanzuala}
 \author{Vasant Natarajan}
 \email{vasant@physics.iisc.ernet.in}
 \homepage{www.physics.iisc.ernet.in/~vasant}
 \affiliation{Department of Physics, Indian Institute of
 Science, Bangalore 560\,012, INDIA}

\begin{abstract}
We have developed a technique to measure the absolute frequencies of optical transitions by using an evacuated Rb-stabilized ring-cavity resonator as a transfer cavity. We study possible wavelength-dependent errors due to dispersion at the cavity mirrors by measuring the frequency of the same transition in the $D_2$ line of Cs at three cavity lengths. We find no discernable change in values within our error of 30~kHz. Our values are consistent with measurements using the frequency-comb technique and have similar accuracy.
\end{abstract}



\maketitle

Precise measurement of the absolute frequencies of atomic transitions have played an important role in the development of physics. This is perhaps best exemplified by the measurement of the small Lamb shift between the $2S_{1/2}$ and $2P_{1/2}$ levels of hydrogen, which led to the birth of quantum electrodynamics (QED), our most-successful physical theory to date. Since the SI standard of frequency or time is defined using a hyperfine transition in the Cs atom ($\approx 9.1$~GHz, which is in the microwave region), it is a challenge to measure the {\it absolute} frequency of an optical transition for it means that one has to span six orders of magnitude without loss in accuracy. In recent times, the use of a femtosecond frequency comb \cite{HAL06} generated by a laser beam in a nonlinear fiber has enabled such a span. The uniform spacing of the comb is in the microwave region and directly referenced to the atomic clock, while the laser frequency is in the visible region.

Around the same time, we have developed an alternate technique to measure the frequencies of optical transitions \cite{BDN03,BDN04,DBB06}, which gives similar accuracy especially when the measurement is limited by errors in spectroscopy. The technique uses an evacuated ring-cavity resonator whose length is calibrated against a hyperfine transition in  the $D_2$ line of Rb. The absolute frequency of this line has been measured to a relative accuracy of $1.4 \times 10^{-11}$ using a frequency chain referenced to the atomic standard \cite{YSJ96}. Thus, the Rb line acts as a secondary standard in the optical regime and the ring-cavity as a transfer cavity, so that the {\it absolute} frequency of an unknown laser can be determined.

\begin{figure}
\centering{\resizebox{0.95\columnwidth}{!}{\includegraphics{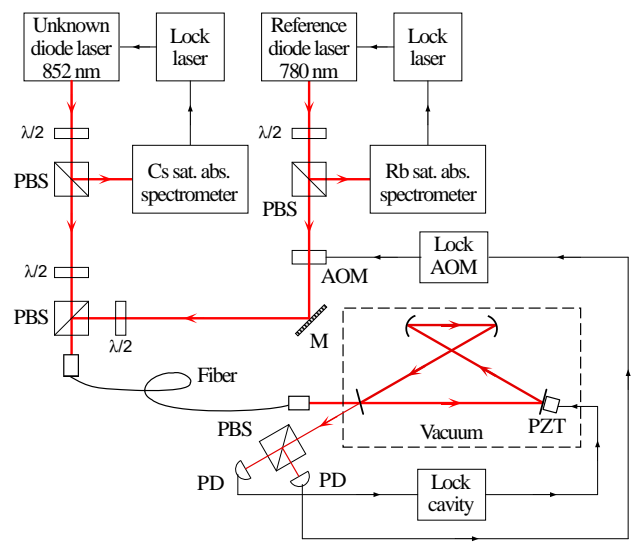}}}
\caption{Schematic of the experiment. Figure key -- $\lambda/2$: halfwave retardation plate, PBS: polarizing beam splitter, AOM: acousto-optic modulator, M: mirror, PD: photodiode, PZT: piezoelectric transducer.}
 \label{schematic}
\end{figure}

The basic idea of the measurement scheme can be understood by referring to Fig.\ \ref{schematic}. The cavity is brought into simultaneous resonance for both lasers, so that the ratio of the two frequencies is a ratio of two {\it integer} mode numbers. While the resonance condition cannot be satisfied simultaneously for two arbitrary frequencies, it can be satisfied if we use an acousto-optic modulator (AOM) to offset the frequency of one of the lasers (the reference laser in our case). The resonance condition for a cavity length $L$ can be written as
\begin{equation}
L = n \lambda \, ,
\label{res1}
\end{equation}
where $n$ is the mode number and $\lambda$ is the wavelength. For an evacuated cavity, $\lambda = c/\nu$, where $\nu$ is the frequency. Thus, the ratio of the unknown to the reference frequencies is given by
\begin{equation}
\frac{\nu_{\rm unk}}{\nu_{\rm ref} + \nu_{\rm AOM}} = \frac{n_{\rm unk}}{n_{\rm ref}}\, .
\label{cav1}
\end{equation}
A knowledge of the cavity length (or mode numbers) combined with a measurement of the AOM frequency constitutes a measurement of the unknown frequency.

Although the use of an evacuated cavity enables us to get the optical frequency, Eq.\ \ref{res1} shows that the technique is essentially a wavelength comparison. We must therefore be careful about possible systematic errors due to wavelength-dependent effects. We have shown earlier that the Guoy phase (due to diffraction) is wavelength {\it independent} in a ring cavity and therefore does not constitute a source of error \cite{DBB06}. Another potential error arises due to wavelength-dependent phase shifts (as a result of dispersion) at the cavity mirrors. This can modify the resonance condition in Eq.\ \ref{res1} either due to a constant phase offset or the Goos-H\"anchen shift, or both. However, we have a good experimental handle to address this source of error, namely to measure the frequency at different cavity lengths. Since the phase error due to dispersion at the mirrors is constant while the accumulated phase due to light propagation increases with length, the values at different cavity lengths should give us an idea of the size of this error.

In this work, we present a comprehensive study of errors due to mirror dispersion by measuring the frequency of the $D_2$ line in Cs at three cavity lengths. In earlier work \cite{DBB06,DAN07}, we have estimated this error by varying the cavity length, but we could vary it only by about 25\%; here we vary it by almost a factor of 2. We find no significant difference in the values over this variation in length. We also compare our values to measurements using a frequency comb, and find reasonably good agreement.

The experimental details are as follows (see Fig.\ \ref{schematic}). The reference laser is a home-built diode laser system which is frequency stabilized using grating feedback \cite{BRW01}. The linewidth after stabilization is about 1~MHz. It is locked to a hyperfine transition on the $D_2$ line of $^{87}$Rb (${5S}_{1/2} \rightarrow {5P}_{3/2}$ transition) at 780~nm using fm modulation in a saturated-absorption spectroscopy set up. The unknown laser is a similar diode laser system operating at 852~nm. It is similarly locked to a hyperfine transition on the $D_2$ line of $^{133}$Cs (${6S}_{1/2} \rightarrow {6P}_{3/2}$ transition), but modulated at a slightly different frequency. The two laser beams with orthogonal linear polarizations are mixed on a polarizing beam splitter (PBS) and coupled into a single-mode fiber. The use of the fiber guarantees that the two beams enter the cavity at exactly the same angle, thereby avoiding potential errors due to differential excitation of higher-order cavity modes. The beams entering the cavity have a size of 2~mm and power of 1~mW each. The cavity is evacuated with a pump and maintained at a pressure of $\sim 10^{-3}$~torr. The power reflected from the input coupler of the cavity is split into its two linear components using a PBS. The second plane mirror of the cavity is mounted on a piezoelectric transducer (PZT). The PZT is modulated at 35~kHz to generate error signals for locking. The demodulated signal from the unknown laser is used to lock the cavity on resonance, while the demodulated signal from the reference laser is used to lock the frequency of the AOM driver. This frequency is read using a frequency counter with a timebase stability of $10^{-8}$.

To establish the cavity length uniquely, the same frequency is measured with the reference laser on an $F=1 \rightarrow F'$ transition, and on an $F=2 \rightarrow F'$ transition. This changes the frequency of the reference laser by a known amount of nearly 6.5~GHz \cite{AIV77}. Then there is only one mode number combination that satisfies Eq.\ \ref{cav1}. In other words, the change of the reference laser frequency by a known amount gives us a measurement of the free-spectral-range (FSR) of the cavity, and hence its length.

The results of measurements at three cavity lengths -- 177, 220, and 315~mm, are listed in Table \ref{t1a}. There are three hyperfine transitions starting from the upper level of the ground state, and we have measured the frequency for each transition. In order to keep statistical errors below 5~kHz, we set the integration time in the frequency counter to 10~s. Then we take an average of 20--30 independent readings. The most important thing to note from the table is that the measured values at the three lengths are within a few kHz of each other. Since the uncertainty of the lock to a cavity peak is about 30~kHz, the observed variation can be explained solely on the basis of lock-point uncertainty. Thus any error due to dispersion at the mirrors is less than our detectable limit. This is also what we had found in earlier studies \cite{DBB06,DAN07}, but there we were only able to reduce the length from 220~mm to 175~mm. Here, we see no significant change even when we increase the length to 315~mm.

\begin{table}
\caption{Transition frequencies measured at three cavity lengths. There are two values at each length, measured with the reference laser locked to either an $F=1 \rightarrow F'$ or an $F=2 \rightarrow F'$ transition. For convenience, a constant offset of 351\,721\,000~MHz has been removed from all values.}
\begin{ruledtabular}
\begin{tabular}{ccccc}
Length & Ref.\ Laser & \multicolumn{3}{c}{Frequency for  $F \rightarrow F'$ (MHz)} \\
\cline{3-5}
\multicolumn{1}{c}{(mm)} & lock point & $4 \rightarrow 3$ & $4 \rightarrow 4$ & $4 \rightarrow 5$ \\
\hline
177 & $1 \rightarrow F'$ & 508.241 & 709.531 & 960.326 \\
\vspace*{2mm}
    & $2 \rightarrow F'$ & 508.229 & 709.524 & 960.350 \\
220 & $1 \rightarrow F'$ & 508.232 & 709.538 & 960.322 \\
\vspace*{2mm}
    & $2 \rightarrow F'$ & 508.244 & 709.514 & 960.388 \\
315 & $1 \rightarrow F'$ & 508.265 & 709.560 & 960.391 \\
    & $2 \rightarrow F'$ & 508.236 & 709.545 & 960.362
\end{tabular}
\end{ruledtabular}
 \label{t1a}
\end{table}

\begin{figure}[t]
\centering{\resizebox{0.5\columnwidth}{!}{\includegraphics{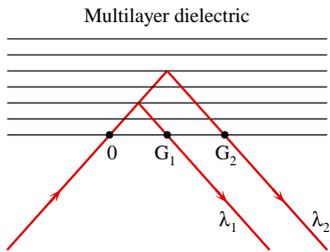}}}
\caption{Goos-H\"anchen shift at a multilayer dielectric mirror for a non-zero angle of incidence. The lateral shift $G_i$ is dependent on the wavelength $\lambda_i$.}
 \label{goos}
\end{figure}

It therefore seems likely that there is some effect that causes the wavelength dependence to cancel in our ring-cavity geometry. One possible explanation is shown in Fig.\ \ref{goos} where we consider the Goos-H\"anchen shift at a multilayer dielectric mirror. The point of emergence ($G_i$) depends on the wavelength ($\lambda_i$). In addition, the phase of the wave at $G_i$ depends on the phase at the reflection point and therefore on $\lambda_i$. It is possible that these two cancel exactly to make the cavity resonance condition wavelength {\it independent}. Note that the Goos-H\"anchen shift will be zero in a linear cavity where the angle of incidence at the mirrors is $0^\circ$, thus there cannot be a cancelation of this kind.

In the above set of measurements, we did not have to worry about spectroscopic errors since we were measuring the same transition at the three lengths. We were only interested in finding out if this value changes, and not the value itself. However, if we can estimate the error in the Cs spectrometer, we can make a comparison of our values to other measurements using frequency combs \cite{URH00,GTD04}. The main error in our spectrum arises because of lineshape distortion due to light pressure and asymmetric optical pumping in the presence of stray magnetic fields. We estimate this error to be of order 30~kHz \cite{DBB06}. Adding this in quadrature with the cavity lock error of 30~kHz, our total error is of about 50~kHz.

The average value for each transition with this error estimate is compared to previous frequency-comb results in Table \ref{t2}. Our values closely overlap with the ones from Udem et al.\ \cite{URH00} for all three transitions. However, the values from Gerginov et al.\ \cite{GTD04} only overlap for the first two transitions. For the third transition (i.e., $F=4 \rightarrow F'=5$), their value is $2 \, \sigma$ away from that of Udem et al. and $4.4 \, \sigma$ away from ours. This transition is also involved in determining the \{$5-4$\} hyperfine interval in the $P_{3/2}$ state. It is perhaps relevant to note that the value of this interval measured by Gerginov et al.\ \cite{GDT03}, presumably using the same spectrometer, is $4.6 \, \sigma$ discrepant from an earlier measurement by Tanner and Wieman \cite{TAW88} and $9 \, \sigma$ discrepant from a measurement from our group \cite{DAN05}.

\begin{table}
\caption{Comparison to previous frequency-comb measurements. For convenience, a constant offset of 351\,721\,000~MHz has been removed from all values.}
\begin{ruledtabular}
\begin{tabular}{llll}
Reference & \multicolumn{3}{c}{Frequency for  $F \rightarrow F'$ (MHz)} \\
\cline{2-4}
& $4 \rightarrow 3$ & $4 \rightarrow 4$ & $4 \rightarrow 5$ \\
\hline
This work & 508.241(50) & 709.535(50) & 960.356(50) \\
Ref.\ \cite{URH00} & 508.225(110) & 709.544(110) & 960.362(110) \\
Ref.\ \cite{GTD04} & 508.211(6) & 709.497(6) & 960.586(6)
\end{tabular}
\end{ruledtabular}
 \label{t2}
\end{table}

Despite this minor discrepancy, we feel that our values are sufficiently consistent with the frequency comb results to be able to conclude that there is no unknown systematic error plaguing our spectrometer. The measured values in Table \ref{t1a} also show that there is no discernible trend due to dispersion when we change the cavity length by almost a factor of two. This is certainly the same observation that we made in earlier studies. We have given a possible explanation for this lack of wavelength dependence in terms of cancelation of the phase shift and the Goos-H\"anchen shift. This is another advantage of the ring cavity over a linear cavity; the Goos-H\"anchen shift is not present in a linear cavity because of the zero angle of incidence on the mirrors.

\section*{Acknowledgments}
This work was supported by the Department of Science and
Technology, India. A.K.S. acknowledges financial support
from the Council of Scientific and Industrial Research,
India.


\end{document}